\documentclass[conference]{IEEEtran}

\usepackage{multirow}
\usepackage{amssymb}
\usepackage[table]{xcolor}
\usepackage{color,soul}
\usepackage{pgfplots}

\usepackage{algorithm}
\PassOptionsToPackage{noend}{algpseudocode}
\usepackage{algpseudocode}
\algnewcommand\And{\textbf{and} }

\usepackage{graphicx}
\ifCLASSINFOpdf
\else
\fi

\begin{document}
%
\title{Towards Railways Remote Driving: Analysis of Video Streaming Latency and Adaptive Rate Control}


\author{\IEEEauthorblockN{Daniel Mej\'ias, Zaloa Fern\'andez, Roberto Viola}
\IEEEauthorblockA{Fundaci\'on Vicomtech\\
Basque Research and Technology Alliance\\
San Sebasti\'an, 20009 Spain\\
Email: \{damejias, zfernandez, rviola\}@vicomtech.org}
\and
\IEEEauthorblockN{Ander Aramburu, Igor Lopez, Andoni Diaz}
\IEEEauthorblockA{Construcciones y Auxiliar de Ferrocarriles I+D (CAF I+D)\\
Beasain, 20200 Spain\\
Email: \{ander.aramburu, igor.lopez, andoni.diaz\}@caf.net}
}


\maketitle

\begin{abstract}
Remote driving aims to improve transport systems by promoting efficiency, sustainability, and accessibility.
In the railway sector, remote driving makes it possible 
to increase flexibility, as the driver no longer has to be in the cab. However, this brings several challenges, as it has to provide at least the same level of safety obtained when the driver is in the cab.
To achieve it, wireless networks and video streaming technologies gain importance as they should provide real-time track visualization and obstacle detection capabilities to the remote driver.
Low latency camera capture, onboard media processing devices, and streaming protocols adapted for wireless links are the necessary enablers to be developed and integrated into the railway infrastructure.
This paper compares video streaming protocols such as Real-Time Streaming
Protocol (RTSP) and Web Real-Time Communication (WebRTC), as they are the main alternatives based on Real-time Transport Protocol (RTP) protocol to enable low latency.
As latency is the main performance metric, this paper also provides a solution to calculate the End-to-End video streaming latency analytically.
Finally, the paper proposes a rate control algorithm to adapt the video stream depending on the network capacity. The objective is to keep the latency as low as possible while avoiding any visual artifacts.
The proposed solutions are tested in different setups and scenarios to prove their effectiveness before the planned field testing.

\end{abstract}


\IEEEpeerreviewmaketitle

\section{Introduction}

The European Union believes that the future of transport lies in dependence on trains, which will require new ways to improve and innovate the railway sector \cite{islam2016make}. Pushing further automation can 
increase safety, and optimize the efficiency of line utilization \cite{trentesaux2018autonomous}.
Thus, remote train driving has become one of the technological enablers for railway companies to achieve this automation. The first advantage is the flexibility of the driver's location. Moreover, it is a viable option for depots, rural areas with difficult access \cite{zieger2021opportunities} or as a rescue mode for autonomous train units not working as expected \cite{goikoetxea2023remote}.

However, developing remote driving capabilities is not trivial, as it is necessary to provide solutions with high reliability and safety as the main objectives. The safety level should be at least the same as the current one with the driver in the cab. Therefore, it is required to establish parameters and rules for automatic operations and safety countermeasures during infrastructure breakdowns or unidentified obstacles \cite{9209333}.

Real-time transmission of data and images becomes crucial, as the latency between the train and the remote driver will depend on the transmission methods and transport protocols employed.
Remote driving also relies on a sufficient network quality for delivering reliable, low-latency video to the remote operator \cite{feng2019toward}. 
Global System for Mobile for Railways (GSM-R), being based on 2G \cite{chen2018development}, is becoming obsolete due to technological advances in mobile communications. In response to new needs, Future Railway Mobile Communication System (FRMCS) has emerged as a 5G system \cite{he20225g}. This technology is based on 3GPP Release 16 (Rel-16).
Rel-16 includes radio enhancements for Ultra Reliable Low Latency Communications (URLLC), while FRMCS defines video quality for railway operations. Even the minimum requirements are H.264 codec with 320x240 frames at 10fps, the recommended ones are 1920x1080 at 30fps to guarantee sufficient quality. Moreover, the video codec shall be adapted in real time to consider the QoS characteristics of the communication channel.
Finally, the driving performance does not depends only on video quality, but also on the train's speed \cite{9797698}.

Real-time Transport Protocol (RTP) provides an end-to-end network transport suitable for video transmission, allowing packet delivery with the shortest delay. Moreover, Real-Time Transport Control Protocol (RTCP) works with RTP to monitor the delivery. The Real-Time Streaming Protocol (RTSP) and Web Real-Time Communication (WebRTC) are on top of RTP/RTCP, describing and initializing the communication between sender and receiver.

This paper proposes solutions for measuring the streaming latency and an onboard media server's rate adaptation. Specifically, this work includes three relevant contributions:
\begin{itemize}
    \item A video streaming architecture for remote driving. It includes industrial cameras and equipment according to railway standards, where a media server generates the video stream to be played from the remote driver's location. RTSP and WebRTC protocols are evaluated on top of this architecture.
    \item Evaluation of the end-to-end video latency by implementing a measurement method compatible with any RTP-based protocol such as RTSP and WebRTC. 
    This method calculates the time elapsed from the capture of the image onboard to the time of displaying at the remote location
    \item An adaptive rate control algorithm for video streaming based on network metrics. RTCP reports are captured at the media server to adjust the video encoding bitrate to the network capacity.
\end{itemize}

The proposed solutions are evaluated in different setups deployed at Vicomtech and CAF laboratories.

This paper is structured as follows. Section \ref{sec:proposed} covers the proposed architecture and methodology for latency measurement and adaptive rate control. Next, in sections \ref{sec:implementation} and \ref{sec:results}, the implementation is explained and the experimental results are analyzed. Finally, section \ref{sec:conclusion} summarizes the results of this research and outlines some open issues.

\section{Architecture and Methodology}
\label{sec:proposed}

\subsection{General architecture}

\begin{figure}[!t]
\centering
\includegraphics[width=0.50\textwidth,clip,keepaspectratio]{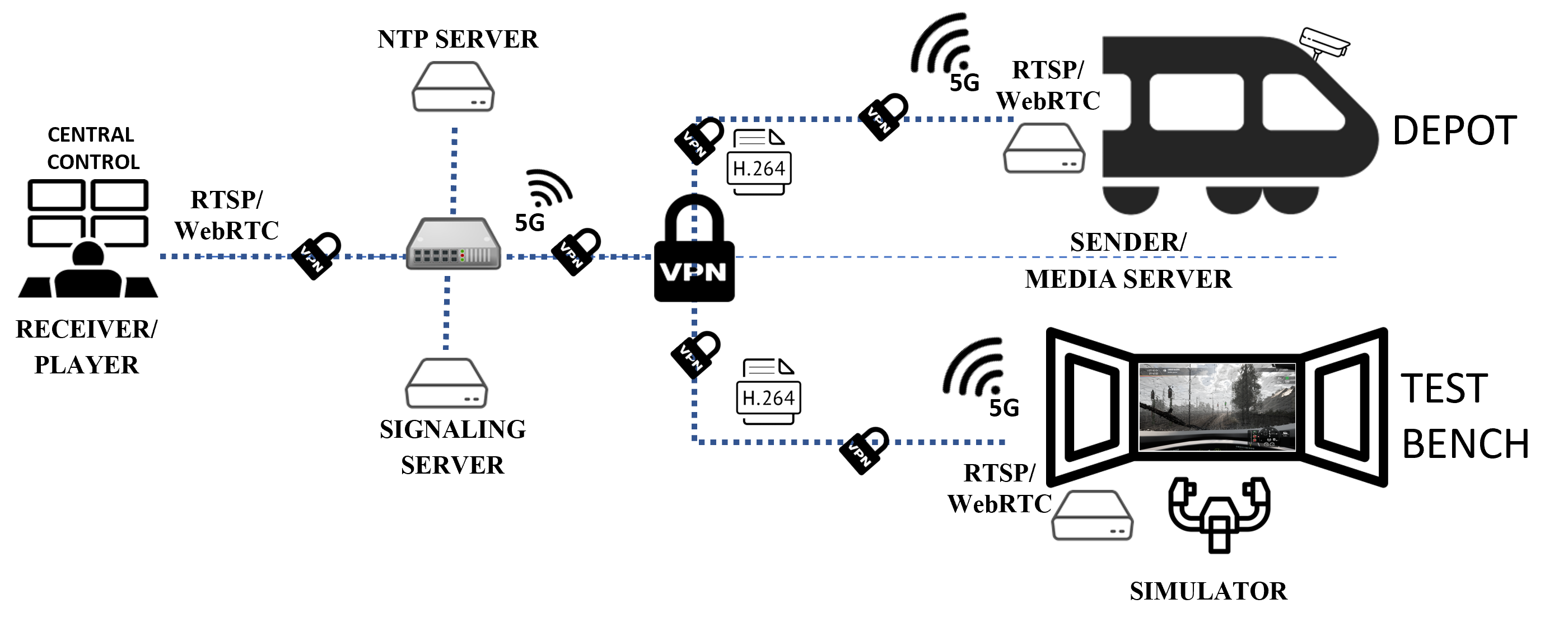}
\caption{5G architecture for remote driving for trains.}
\label{fig:diagram_general}
\end{figure}

Figure \ref{fig:diagram_general} shows the general architecture of the proposed solutions. This architecture considers both test and real scenarios. It allows the later installation of the tested solutions on a rail vehicle so that the transition from the test environment to the real environment is natural and simple.

In both test and real cases, there is sending and receiving equipment. The receiver consists of a player that allows the remote driver to visualize the video captured on the train. The sender embeds a media server responsible for capturing video from the cameras and transmitting it to the remote player via RTSP or WebRTC. At the same time, a Network Time Protocol (NTP) server is needed to synchronize both sender and receiver. A signaling server is required for WebRTC negotiation only. Communication between the sender and receiver is based on a 5G Standalone (SA) network, while a Virtual Private Network (VPN) ensures compliance with security requirements.

\subsection{End-to-End Latency measurements}

\begin{figure}[!t]
\centering
\includegraphics[width=0.45\textwidth,clip,keepaspectratio]{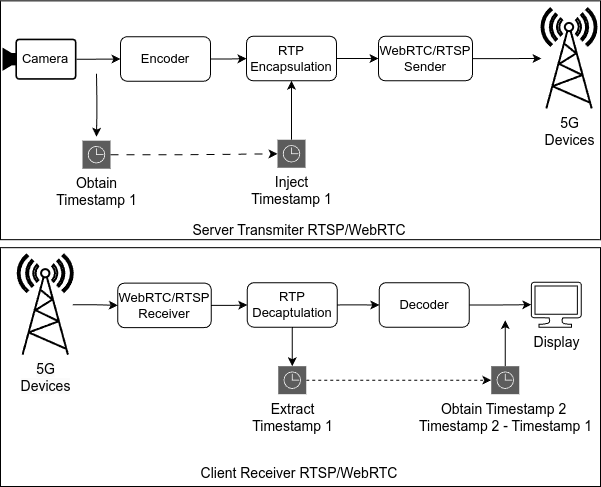}
\caption{Camara Basler RTSP/WebRTC transport diagram.}
\label{fig:Timestamp_Inject}
\end{figure}

The solution for the End-to-End latency measurement is depicted in Figure \ref{fig:Timestamp_Inject}. This calculation requires operations at the onboard media server and the player. The server obtains the timestamp when capturing the image (\textit{Timestamp1}) and adds it to the RTP packets generated after the encoding. Then, the player retrieves the timestamp from the RTP packets and compares it with the current time (\textit{Timestamp2}) when the image is displayed (\textit{Timestamp2}-\textit{Timestamp1}).

The NTP synchronization is important to guarantee accurate timestamps. Moreover, it is proposed to transmit a 64-bit timestamp over RTP for higher accuracy, as it includes nanosecond resolution.
The 64-bit timestamp is divided into two 32-bit blocks. The RTP header has 32 bits meant for timestamps where the first block is added, while RFC 8286 defines an RTP extension header that allows storing the second block of 32 bits \cite{deng2017internet}. The use of RTP header and its extension guarantee compatibility with both RTSP and WebRTC.

Going deeper, the process carried out at the media server is the following:
\begin{itemize}
    \item Capture: the camera captures an image together with the timestamp. The timestamp is added to the metadata of the image.
    \item Encoding: the image is encoded into a
    H.264 bitstream \cite{wiegand2003overview}. The metadata is maintained unaltered along the encoding process.
    \item Encapsulation: the H.264 video stream is encapsulated into RTP payload \cite{wenger2005rfc}. The capture timestamp is extracted from the metadata and added to the RTP header. For this, it is required both the RTP standard header and its RFC 8286 extension.
    \item Sender: RTP packets are sent on the communication channel. In the case of RTSP, the player opens a connection with the sender. For WebRTC, a negotiation between the sender and receiver is performed through the signaling server to determine the communication route.
\end{itemize}

The player receives the RTP packets through RTSP or WebRTC and calculates the latency:
\begin{itemize}
    \item Receiver: it receives the RTP packets through the channel established with the media server.
    \item Decapsulation: the original H.264 content is extracted from the RTP payload. In addition, the timestamp contained in the RTP header is extracted and added as metadata of the H.264 content.
    \item Decoding: the H.264 content is decoded to retrieve the uncompressed image. The metadata is maintained unaltered along the decoding process. 
    \item Displaying: the image is displayed. Moreover, the timestamp is extracted from the metadata and subtracted from the current time to obtain the End-to-End latency. This is shown to the remote driver, who will consider it during the operations.
\end{itemize}

\subsection{Adaptative Rate control}

Algorithm \ref{alg:algorithm} describes in detail our proposed bitrate adaptation mechanism executed at the sender. Accordingly, two decision factors and three different quality levels are considered.

The jitter (jit) is the quantitative variation in data arrival time, measured in units of the system clock frequency. 
The fraction of packet loss (fpl) is the packet loss expressed as a fraction of the total packets sent or received.

The jitter and fraction of packet loss are obtained at the server from the information contained in the RTCP report received from the client (player).
The optimal bitrate is chosen when the fraction of packet loss and jitter values are good enough.
The bitrate is reduced when packet loss or jitter is not optimal. When packet loss is detected or the jitter level exceeds the good threshold (fpl$^{h,m}$, jit$^{h,m}$), the bitrate is reduced to recover the transmission stability. If the jitter threshold exceeds the medium level and packet losses are detected (fpl$^{m,l}$, jit$^{m,l}$), the bitrate is adjusted to the low level to maintain stability. For the opposite case, if the jitter is below the medium level and there is no packet loss, the bitrate is increased to the medium level to improve the image quality. If the jitter level is below the good threshold and there is no packet loss, the bitrate is increased to a good level for optimal image quality.
In this way, adequate video transmission is provided to allow the remote operator to visualize in detail the obstacles when network bandwidth is observed to be decreasing.

\begin{algorithm}
\renewcommand{\algorithmicrequire}{\textbf{Input:}}
\renewcommand{\algorithmicensure}{\textbf{Output:}}
\caption{3-levels Adaptive Rate Control}
\label{alg:algorithm}
\begin{algorithmic}
\Function{adaptiveRate}{}
\Comment{\parbox[t]{0.47\linewidth}{when a RTCP report comes}}
\Require bitrate$_{t}$ \Comment{current bitrate}
\Require [bitrate$^{high}$,bitrate$^{medium}$,bitrate$^{low}$] \Comment{\parbox[t]{0.2\linewidth}{admissible bitrates}}
\Require [fpl$^{h,m}$, jit$^{h,m}$] \Comment{threshold high-medium}
\Require [fpl$^{m,l}$, jit$^{m,l}$] \Comment{threshold medium-low}
\Require fpl$_{t}$, jit$_t$ \Comment{measured packets loss fraction and jitter}
\Ensure bitrate$_{t+1}$ \Comment{next bitrate}
\If {(bitrate$_{t}$) $=$  (bitrate$^{high}$)}
    \If {(fpl$_{t}$, jit$_{t}$) $<$  (fpl$^{h,m}$, jit$^{h,m}$)}
        \State bitrate$_{t+1}$ $\leftarrow$ bitrate$^{high}$ \Comment{high bitrate}
    \Else 
        \State bitrate$_{t+1}$ $\leftarrow$ bitrate$^{medium}$ \Comment{medium bitrate}
    \EndIf
\ElsIf {(bitrate$_{t}$) $=$  (bitrate$^{medium}$)}
    \If {(fpl$_{t}$, jit$_{t}$) $<$  (fpl$^{h,m}$, jit$^{h,m}$)}
        \State bitrate$_{t+1}$ $\leftarrow$ bitrate$^{high}$ \Comment{high bitrate}
    \ElsIf {(fpl$_{t}$, jit$_{t}$) $>$  (fpl$^{m,l}$, jit$^{m,l}$)}
        \State bitrate$_{t+1}$ $\leftarrow$ bitrate$^{low}$ \Comment{low bitrate}
    \Else
        \State bitrate$_{t+1}$ $\leftarrow$ bitrate$^{medium}$ \Comment{medium bitrate}
    \EndIf
\ElsIf {(bitrate$_{t}$) $=$  (bitrate$^{low}$)}
    \If {(fpl$_{t}$, jit$_{t}$) $<$  (fpl$^{m,l}$, jit$^{m,l}$)}
        \State bitrate$_{t+1}$ $\leftarrow$ bitrate$^{medium}$ \Comment{medium bitrate}
    \Else
        \State bitrate$_{t+1}$ $\leftarrow$ bitrate$^{low}$ \Comment{low bitrate}
    \EndIf
\EndIf
\EndFunction
\end{algorithmic}
\end{algorithm}

\section{Implementation}
\label{sec:implementation}

For the development of the proposed method, the GStreamer framework \cite{Gstreamer} has been used. Specifically, the following elements are employed:

\begin{itemize}
    \item Pylon source: the official Basler element that allows the capture of camera images and timestamps.
    \item H.264 NVidia encoder/decoder: elements provided by NVIDIA for its graphics cards. The encoder configuration is key to enable bitrate adaptation based on measured network statistics.
    \item RTP H.264 pay/depay: this packages the H.264 encoded frames into RTP packets \cite{wenger2005rfc}. RTP packets include the camera's timestamp in their headers.
    \item WebRTCbin: it allows communication via WebRTC, enabling the peer-to-peer connection for sending and receiving video. It must connect to the signaling server, which is responsible for negotiation between the peers.
    \item RTSP server/client: they allow the creation of an RTSP server to manage connections and send data, and a client to initiate a connection and receive data.

\end{itemize}

The video to be transmitted must be optimal for transmitting content with maximum sharpness, detail and visibility. It must be able to visualize the image with the highest precision and detail so that the operator can differentiate each object and obstacle on the rails. For this purpose, the video profiles shown in the table \ref{tab:quality} have been defined. These profiles allow for high quality and will be adjusted according to the degradation of the transmission channel.

\begin{table}[!t]
    \caption{Video profiles}
    \centering
    \begin{tabular}{|c|c|c|c|c|}
    \hline
        Quality & Resolution & Codec & Bitrate & Framerate  \\
    \hline
         HIGH & 1920x1080 & H.264 & 5Mbps & 30fps \\
         MEDIUM & 1920x1080 & H.264 & 3.5Mbps & 30fps \\
         LOW & 1920x1080 & H.264 & 2Mbps & 30fps \\
    \hline
    \end{tabular}
    \label{tab:quality}
\end{table}

The adaptive rate control parameters for the signal degradation will be defined by a threshold as shown in table \ref{tab:threshold}. Two thresholds are defined. High-Medium is the threshold for changing quality between High and Medium levels, while Medium-Low is meant between Medium and Low levels.

\begin{table}[!t]
    \caption{Thresholds for RTCP metrics.}
    \centering
    \begin{tabular}{|c|c|c|}
    \hline
        Threshold & Packet Loss Fraction (\%) & Jitter (Hz)  \\
    \hline
         High-Medium &  2 & 500 \\
         Medium-Low & 2 & 1000 \\
    \hline
    \end{tabular}
    \label{tab:threshold}
\end{table}

Two laboratories will be used for the testing of the solution. These laboratory deployments are prior steps to a future outdoor one on a real train cabin and remote control center. Each laboratory has different network characteristics to investigate different aspects of the solution.

\subsection{Vicomtech laboratory setup}

\begin{figure}[htp]
\centering
\includegraphics[width=0.45\textwidth,clip,keepaspectratio]{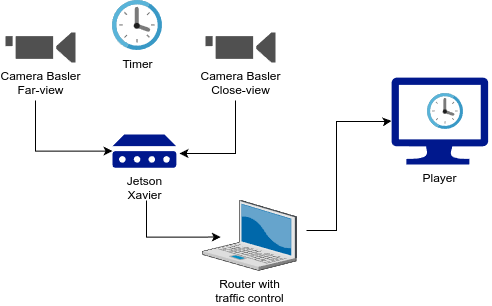}
\caption{Vicomtech setup diagram.}
\label{fig:Vicomtech_Setup}
\end{figure}

In the Vicomtech laboratory, an architecture was deployed that follows the configuration for a train cabin. The test approach would have two cameras, onboard equipment and player equipment for visualization as shown in Figure \ref{fig:Vicomtech_Setup}. For these tests, portable equipment with router simulation and traffic control will be used to evaluate the End-to-End latency measurement method and the adaptive rate control method.

\begin{itemize}
    \item Cameras: Two Basler cameras model acA1920-40uc are used, one of them with focus to detail objects and obstacles, long distance focus is essential(far-view). Another camera is used for close-up viewing of roads and obstacles with short distance visual focus(close-view).
    \item Media Server: a server deployed on a Jetson Xavier with ARM processor and image processing elements for encoder and decoder of 4K images at 30fps and H.264 encoding have been used. A WebRTC and RTSP server is deployed for each corresponding test.
    \item Network equipment: a switch and a laptop simulating a router have been used for the communication between the server RTSP and player. Using a laptop as a router will allow us to perform tests with network attenuation to evaluate the behavior of the applications against bandwidth degradation.
    \item Laptop (Player): equipment with the developed receiver application that allows measuring the End-to-End latency and transmits the necessary data to the server for the adaptation method. A second laptop was used to contain the player and to differentiate it from the laptop used as a router.
\end{itemize}

\subsection{CAF laboratory setup}

\begin{figure}[htp]
\centering
\includegraphics[width=0.45\textwidth,clip,keepaspectratio]{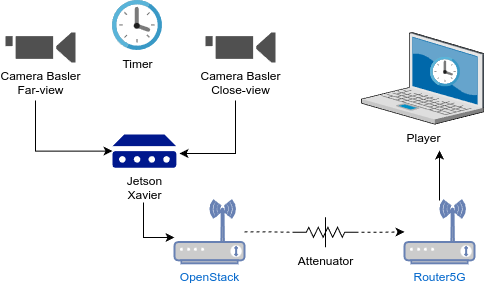}
\caption{CAF setup diagram.}
\label{fig:Beasain_Setup}
\end{figure}

For the CAF laboratory, equipment was deployed to simulate the transmission of one of the train cabins and used a 5G network with an attenuator as shown in Figure \ref{fig:Beasain_Setup}. The attenuator simulates the attenuation that can be found in the environment. In addition, the operator's player will have different visualization software in order to make a comparison of the best performing software. The comparison is made between the one developed during this project and the software available from open source. 

This deployment will be replicated in field tests, where the equipment will be installed in the cabs of a train, and the remote driving tests will be performed in a train test field. The equipment to be installed must meet railway standards, as it must be able to withstand conditions, stabilities, and safety standards that are present in railway vehicles.  

The CAF laboratory deployment focused on system behavior in the 5G network. Screen-to-screen latency was evaluated in order to obtain accurate measurements. The equipment that will be present in the cabin are the following:

\begin{itemize}
    \item Cameras: There will be two cameras in each cabin for far-view and close-view. Basler cameras model acA1920-40uc will be used that provide low latency transmission with high visual accuracy. 
    \item Onboard server: A Jetson Xavier with specifications for railroad standards will be used for this equipment. This equipment will be in charge of receiving the images from the cameras and sending the images over the RTSP application. 
    \item Operator: Computer containing the player to visualize video from both cameras.
    \item Router/Switch: For network communication, network equipment will be used that will comply with railroad standards and will be capable of transmitting to the base station using 5G communications. In addition, to comply with information security, the communication between the different equipment will be done with a VPN.
    \item Amarisoft: it is a PC-based equipment solution ideal for supporting 5G functional and performance testing. The equipment runs on a deployment-level 5G software package that offers the same level of baseband functionality as an indoor/outdoor network.
    \item Attenuator: A model RCDAT-6000-110 variable attenuation element will be used to adjust the bandwidth parameters in the 5G transmission channel. The parameters used were an attenuation range of 0 - 110 dB with a step of 0.25 dB for frequencies from 1 to 6000 MHz.
\end{itemize}

\section{Results and Discussion}
\label{sec:results}

Table \ref{tab:rtsp1080} shows the tests done in the Vicomtech laboratory with a two-camera configuration and RTSP. In these tests, the same bitrate has been maintained at 5 Mbps (mode static) for both cameras and the network bandwidth has been reduced.
The obtained results compare the latency from the time after capture to the time before display on the player (End-to-End or E2E) versus the latency from the time in front of the camera to the time displayed on the player image (Screen-to-Screen or S2S).
For these tests, it can be concluded that the transmission latency is approximately 150 ms in S2S when there is enough bandwidth to transmit both videos
In contrast, in the E2E measurements, the obtained latency is 60-70 ms. This is expected since E2E cannot consider the time it takes for the camera to capture the images and the time it takes to represent the decoded video (raw bits) on the screen.

When two videos are transmitted at 5 Mbps each, 
it is observed that from 10 Mbps the traffic starts to compete for the medium, increasing the latency. From 7 Mbps, one of the cameras starts to pixelate, and at 5 Mbps, it is no longer possible to receive video from both cameras, and the image freezes. 

\begin{table}[!t]
\caption{Two cameras Latency in static mode to RTSP/1080p/30fps/5Mbps.
}
\centering
\bgroup
\def\arraystretch{1.2}
\setlength\tabcolsep{2.5pt} 
\label{tab:rtsp1080}
\begin{tabular}{|c|c|c|c|c|}
\hline
\multirow{2}{*}{\textbf{\shortstack{Limit Bandwitdth\\ (Mbps)}}} &
    \multicolumn{2}{|c|}{\textbf{Far-view (ms)}} & \multicolumn{2}{|c|}{\textbf{Close-view (ms)}} \\  \cline{2-5}
 & \textbf{S2S} & \textbf{E2E} & \textbf{S2S} & \textbf{E2E} \\ 
\hline
12	&  149.82&	72.28&  149.82& 60.90\\
11	&  149.83&	75.00&  149.83& 62.99\\
10	&  567.14&	466.42&  567.14& 461.58\\
9	&  1282.84&	1208.89&  1278.70& 1215.84\\
8	&  1416.08&	1356.27&  1449.37& 1369.74\\
7	&  Pixelation &	1559.56&  1614.72& 1528.54\\
6	&  Pixelation& 1777.86&  1865.24& 1736.59\\
5	&  Freezing& Freezing&  Freezing& Freezing\\
4	&  Freezing&	Freezing&  Freezing& Freezing\\
\hline
\end{tabular}
\egroup
\end{table}

Table \ref{tab:webrtc1080} shows the results of the test performed in the Vicomtech Laboratory obtained with the same configuration, in static mode and latency measurement as in the RTSP case shown in Table \ref{tab:rtsp1080}, but using WebRTC. The S2S latency measurements remain in the range of 150 ms and for E2E in 60 ms, obtaining similar results to RTSP. As with RTSP, the video starts to pixelate after 7 Mbps due to both cameras competing for the channel. 

WebRTC has the advantage of being able to facilitate point-to-point communication but may be less suitable when multiple simultaneous connections to a single source, such as a camera, are required. In contrast, RTSP offers flexibility and extended control over the media stream. This is why the rest of the tests conducted have been conducted with RTSP, as multiple access to the camera is necessary for remote control in trains. It might be viewed from the cab for testing purposes and from the control center for remote viewing.

\begin{table}[!t]
\caption{Two cameras Latency in static mode to WebRTC/1080p/30fps/5Mbps.
}
\centering
\bgroup
\def\arraystretch{1.2}
\setlength\tabcolsep{2.5pt} 
\label{tab:webrtc1080}
\begin{tabular}{|c|c|c|c|c|}
\hline
\multirow{2}{*}{\textbf{\shortstack{Limit Bandwitdth\\ (Mbps)}}} &
    \multicolumn{2}{|c|}{\textbf{Far-view (ms)}} & \multicolumn{2}{|c|}{\textbf{Close-view (ms)}} \\  \cline{2-5}
 & \textbf{S2S} & \textbf{E2E} & \textbf{S2S} & \textbf{E2E} \\ 
\hline
12	&  150.19&	44.82&  133.45& 45.51\\
11	&  150.24&	55.12&  167.08& 63.62\\
10	&  516.79&	405.63&  500.02& 437.07\\
9	&  1331.51&	1208.90&  1315.32& 1199.96\\
8	&  Pixelation&	1381.79&  Pixelation& 1347.75\\
7	&  Pixelation&	1560.26&  Pixelation& 1551.39\\
6	&  Pixelation& 1796.62&  Pixelation& 1795.16\\
5	&  Freezing& Freezing&  Freezing& Freezing\\
4	&  Freezing&	Freezing&  Freezing& Freezing\\
\hline
\end{tabular}
\egroup
\end{table}

Table \ref{tab:degrade_table} shows the results of the test performed in the Vicomtech Laboratory obtained in the measurements performed by applying the rate control adaptation method in RTSP. The adaptation is performed with the previously defined profiles. It can be observed that the adaptation helps to maintain stability at a lower bandwidth than the static mode since the video is transmitted at a lower bitrate. When the bandwidth is lower than the bitrate of the lowest profile, the video starts to pixelate and freeze. Then, when the bandwidth increases, the video improves quality and provides more bitrate.

Although the video transmission is maintained, the latency is affected. The latency decreases when performing profile changes in the adaptive mode, although the channel remains saturated and does not adjust to the optimal values. Adjusting the changes before the channel becomes saturated can be ideal, thus maintaining a constant flow without the cameras competing for the medium. Similarly, in ideal cases, the video stream maintains a latency of 150 ms S2S and 60-70 ms E2E.

\begin{figure}[!t]
\centering
\includegraphics[width=0.5\textwidth,clip,keepaspectratio]{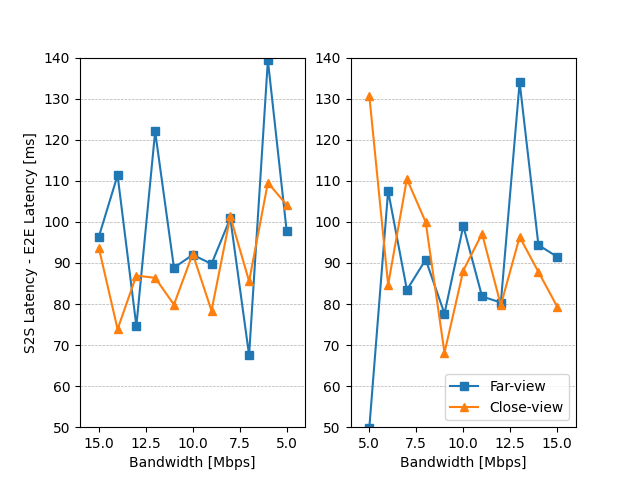}
    \caption{Difference between S2S Latency and E2E Latency when the bandwidth decreases (left) and increases (right)}
    \label{fig:difference}
\end{figure}

To evaluate the time difference between E2E and S2S, Figure \ref{fig:difference} represents the subtraction between them in the adaptive mode test results. On average, a difference of 90 ms between S2S and E2E is obtained, representing the latency added to the image capture by the camera plus the visual display on the screen. This time cannot be obtained directly as it depends on the camera Basler hardware and the monitor on which it is displayed. For this reason, it is necessary to perform the calculation with the S2S comparison.

\begin{table}[!t]

\centering
\bgroup
\def\arraystretch{1.2}
\setlength\tabcolsep{2.5pt} 
\caption{Two cameras Latency in Adaptive mode to RTSP/1080p/30fps.\\
\textcolor[HTML]{bbe33d}{$\blacksquare$} High 
\textcolor[HTML]{e6e905}{$\blacksquare$} Medium 
\textcolor[HTML]{ffb66c}{$\blacksquare$} Low 
\textcolor[HTML]{ffa6a6}{$\blacksquare$} Freezing
}
\label{tab:degrade_table}
\begin{tabular}{|c|c|c|c|c|}
\hline
\multirow{2}{*}{\textbf{\shortstack{Limit Bandwitdth\\ (Mbps)}}} &\multicolumn{2}{|c|}{\textbf{Far-view (ms)}} & \multicolumn{2}{|c|}{\textbf{Close-view (ms)}} \\  \cline{2-5}

 &\textbf{S2S} & \textbf{E2E} & \textbf{S2S} & \textbf{E2E} \\ 
\hline
15 &\cellcolor[HTML]{bbe33d} 148.90& \cellcolor[HTML]{bbe33d} 52.55& \cellcolor[HTML]{bbe33d} 148.90& \cellcolor[HTML]{bbe33d} 55.20\\
14 &\cellcolor[HTML]{bbe33d} 165.51& \cellcolor[HTML]{bbe33d} 54.16& \cellcolor[HTML]{bbe33d} 165.51& \cellcolor[HTML]{bbe33d} 91.63\\
13 &\cellcolor[HTML]{bbe33d} 133.00& \cellcolor[HTML]{bbe33d} 58.36& \cellcolor[HTML]{bbe33d} 167.07& \cellcolor[HTML]{bbe33d} 80.15\\
12 &\cellcolor[HTML]{bbe33d} 185.39& \cellcolor[HTML]{bbe33d} 63.38& \cellcolor[HTML]{bbe33d} 185.39& \cellcolor[HTML]{bbe33d} 99.02\\
11 &\cellcolor[HTML]{bbe33d} 281.59& \cellcolor[HTML]{bbe33d} 192.76& \cellcolor[HTML]{bbe33d} 281.59& \cellcolor[HTML]{bbe33d} 201.73\\
10 &\cellcolor[HTML]{bbe33d} 849.74& \cellcolor[HTML]{bbe33d} 757.76& \cellcolor[HTML]{e6e905}  950.53& \cellcolor[HTML]{e6e905} 858.30\\
9 &\cellcolor[HTML]{bbe33d} 917.22& \cellcolor[HTML]{bbe33d} 827.46& \cellcolor[HTML]{e6e905}  884.66& \cellcolor[HTML]{e6e905} 806.31\\
8	&\cellcolor[HTML]{e6e905}  616.22& \cellcolor[HTML]{e6e905} 515.34& \cellcolor[HTML]{ffb66c} 616.22& \cellcolor[HTML]{ffb66c} 514.88\\
7 &\cellcolor[HTML]{e6e905}  698.10& \cellcolor[HTML]{e6e905} 630.57& \cellcolor[HTML]{e6e905}  729.74& \cellcolor[HTML]{e6e905} 644.20\\
6	&\cellcolor[HTML]{e6e905}  1815.29& \cellcolor[HTML]{e6e905} 1675.94& \cellcolor[HTML]{ffb66c} 1815.29& \cellcolor[HTML]{ffb66c} 1705.74\\
5 &\cellcolor[HTML]{ffb66c} 2059.21& \cellcolor[HTML]{ffb66c} 1961.42& \cellcolor[HTML]{ffb66c} 2097.05& \cellcolor[HTML]{ffb66c} 1992.92\\
4	&\cellcolor[HTML]{ffb66c} Pixelation& \cellcolor[HTML]{ffb66c} 2439.48& \cellcolor[HTML]{ffb66c} Pixelation& \cellcolor[HTML]{ffb66c} 2441.39\\
3 &\cellcolor[HTML]{ffb66c} Pixelation& \cellcolor[HTML]{ffb66c} 3261.87& \cellcolor[HTML]{ffb66c} Pixelation& \cellcolor[HTML]{ffb66c} 3178.43\\
2 &\cellcolor[HTML]{ffa6a6}  Freezing & \cellcolor[HTML]{ffa6a6}  Freezing & \cellcolor[HTML]{ffa6a6}  Freezing & \cellcolor[HTML]{ffa6a6}  Freezing \\
3 &\cellcolor[HTML]{ffa6a6}  Freezing & \cellcolor[HTML]{ffa6a6}  Freezing & \cellcolor[HTML]{ffa6a6}  Freezing & \cellcolor[HTML]{ffa6a6}  Freezing \\
4	&\cellcolor[HTML]{ffb66c} Pixelation& \cellcolor[HTML]{ffb66c} 2605.89& \cellcolor[HTML]{ffb66c} Pixelation& \cellcolor[HTML]{ffb66c} 2755.82 \\
5 &\cellcolor[HTML]{ffb66c} 2082.25& \cellcolor[HTML]{ffb66c} 2032.35& \cellcolor[HTML]{ffb66c} 2247.92& \cellcolor[HTML]{ffb66c} 2117.25 \\
6	&\cellcolor[HTML]{ffb66c} 1980.76&  \cellcolor[HTML]{ffb66c} 1873.21& \cellcolor[HTML]{ffb66c} 2013.58& \cellcolor[HTML]{ffb66c} 1928.89 \\
7 &\cellcolor[HTML]{e6e905}  1649.37& \cellcolor[HTML]{e6e905} 1565.94& \cellcolor[HTML]{ffb66c} 1448.84& \cellcolor[HTML]{ffb66c} 1338.38 \\
8	&\cellcolor[HTML]{e6e905}  283.29& \cellcolor[HTML]{e6e905} 192.48& \cellcolor[HTML]{e6e905}  250.36& \cellcolor[HTML]{e6e905} 150.43 \\
9 &\cellcolor[HTML]{e6e905}  149.26& \cellcolor[HTML]{e6e905} 71.61& \cellcolor[HTML]{e6e905}  149.26& \cellcolor[HTML]{e6e905} 81.14 \\
10	&\cellcolor[HTML]{e6e905}  247.54& \cellcolor[HTML]{e6e905} 148.55& \cellcolor[HTML]{bbe33d} 184.17 & \cellcolor[HTML]{bbe33d} 96.04 \\
11 &\cellcolor[HTML]{bbe33d} 782.13& \cellcolor[HTML]{bbe33d} 700.30& \cellcolor[HTML]{bbe33d} 816.01 & \cellcolor[HTML]{bbe33d} 718.88 \\
12	&\cellcolor[HTML]{bbe33d} 500.68& \cellcolor[HTML]{bbe33d} 420.32& \cellcolor[HTML]{bbe33d} 500.68 & \cellcolor[HTML]{bbe33d} 420.95 \\
13	&\cellcolor[HTML]{bbe33d} 184.08& \cellcolor[HTML]{bbe33d} 50.08& \cellcolor[HTML]{bbe33d} 184.08 & \cellcolor[HTML]{bbe33d} 87.74 \\
14	&\cellcolor[HTML]{bbe33d} 149.23& \cellcolor[HTML]{bbe33d} 54.87& \cellcolor[HTML]{bbe33d} 149.23 & \cellcolor[HTML]{bbe33d} 61.41 \\
15	&\cellcolor[HTML]{bbe33d} 149.69& \cellcolor[HTML]{bbe33d} 58.20& \cellcolor[HTML]{bbe33d} 149.69 & \cellcolor[HTML]{bbe33d} 70.33 \\
\hline
\end{tabular}

\egroup
\end{table}

In the CAF laboratory, the system tests were performed with a 5G network as shown in Table \ref{tab:beasain1080}. 
For this, Amarisoft equipment has been used to deploy the 5G SA network and an attenuator to simulate the degradation of the 5G network. 
In this test, as the real value of the transmission latency is sought, only S2S was measured. The adaptive rate control method could not be implemented because the system degraded very fast at high attenuations.
The results show that with attenuations up to 55 dB of the 5G network, a bandwidth of 18 Mbps is obtained, maintaining a stable network for the transfer of the Far-view and Close-view cameras and obtaining a latency of approximately 200 ms. Already exceeding 60dB, the network is very limited, where communication with the Jetson server is already lost, obtaining a latency of 2 seconds, and after a while, the image freezes. 

Further tests will be executed after installing the solution on a real train at CAF facility in Zaragoza, Spain, and in an outdoor environment in Oslo, Norway. These will have the integration of further elements, such as an improved remote player with the capability to send commands to the train.

\begin{table}[!t]
\caption{Latency in static mode to RTSP/1080p/30fps/5Mbps.
}
\centering
\bgroup
\def\arraystretch{1.2}
\setlength\tabcolsep{2.5pt} 
\label{tab:beasain1080}
\begin{tabular}{|c|c|c|c|}
\hline
\multirow{2}{*}{\textbf{\shortstack{Attenuation \\ (dB)}}} & \multirow{2}{*}{\textbf{\shortstack{Measured Bandwitdth\\ (Mbps)}}} &
    \textbf{Far-view (ms)} & \textbf{Close-view (ms)} \\  \cline{3-4}
 & & \textbf{S2S} & \textbf{S2S} \\ 
\hline
10	& 53.2 & 182.935& 182.935\\
20	& 53.3 & 165.47& 165.47\\
30	& 43.3 & 165.804& 166.004\\
40	& 39.5 & 182.751& 182.751\\
45	& 31.7 & 177.501& 183.616\\
50	& 21.7 & 166.31& 200.31\\
55	& 18 & 199.503& 216.687\\
60	& connection lost & 2731.297& 2730.854\\
\hline
\end{tabular}
\egroup
\end{table}

\section{Conclusions and Future Work}
\label{sec:conclusion}

This paper proposes an End-to-End latency measurement method that can be applied to any streaming protocol based on RTP, such as RTSP and WebRTC. In addition, an adaptive rate control is implemented to adjust the bitrate according to the network state to achieve low latency performance. These solutions are proposed for train remote driving, where low-latency video transmission is required. High latency or network disconnection implies a loss of remote operational availability.

The solutions are tested in a wired network and in a real 5G network, where attenuation is added to the network in both scenarios to simulate the real outdoor behavior. The network has been limited to evaluate the proposed methods' performance.

In the future, it is expected to conclude the equipment installation in a train and perform the outdoor tests in a real remote driving scenario. Moreover, further improvements could address two main aspects. First, the adaptive rate control could also consider the video resolution. In this context, some limitations are envisioned if hardware encoders do not allow to adjust it in real time. Second, the network metrics from RTCP reports could be improved by enabling congestion control algorithms that aim at estimating the available network bandwidth.

\section*{Acknowledgment}

This work was developed within the R2Dato research project inside the EUrail framework (grant agreement No. 101102001). The authors would like to thank Mikel Labayen and Jon Mikel Olmos from CAF Signalling for their support.

\bibliographystyle{IEEEtran}
\bibliography{main.bib}

\begin{thebibliography}{10}
\providecommand{\url}[1]{#1}
\csname url@samestyle\endcsname
\providecommand{\newblock}{\relax}
\providecommand{\bibinfo}[2]{#2}
\providecommand{\BIBentrySTDinterwordspacing}{\spaceskip=0pt\relax}
\providecommand{\BIBentryALTinterwordstretchfactor}{4}
\providecommand{\BIBentryALTinterwordspacing}{\spaceskip=\fontdimen2\font plus
\BIBentryALTinterwordstretchfactor\fontdimen3\font minus \fontdimen4\font\relax}
\providecommand{\BIBforeignlanguage}[2]{{%
\expandafter\ifx\csname l@#1\endcsname\relax
\typeout{** WARNING: IEEEtran.bst: No hyphenation pattern has been}%
\typeout{** loaded for the language `#1'. Using the pattern for}%
\typeout{** the default language instead.}%
\else
\language=\csname l@#1\endcsname
\fi
#2}}
\providecommand{\BIBdecl}{\relax}
\BIBdecl

\bibitem{islam2016make}
D.~M.~Z. Islam, S.~Ricci, and B.-L. Nelldal, ``How to make modal shift from road to rail possible in the european transport market, as aspired to in the eu transport white paper 2011,'' \emph{European transport research review}, vol.~8, no.~3, pp. 1--14, 2016.

\bibitem{trentesaux2018autonomous}
D.~Trentesaux, R.~Dahyot, A.~Ouedraogo, D.~Arenas, S.~Lefebvre, W.~Sch{\"o}n, B.~Lussier, and H.~Ch{\'e}ritel, ``The autonomous train,'' in \emph{2018 13th Annual Conference on System of Systems Engineering (SoSE)}.\hskip 1em plus 0.5em minus 0.4em\relax IEEE, 2018, pp. 514--520.

\bibitem{zieger2021opportunities}
S.~Zieger and N.~Niessen, ``Opportunities and challenges for the demand-responsive transport using highly automated and autonomous rail units in rural areas,'' in \emph{2021 IEEE Intelligent Vehicles Symposium (IV)}.\hskip 1em plus 0.5em minus 0.4em\relax IEEE, 2021, pp. 77--82.

\bibitem{goikoetxea2023remote}
J.~Goikoetxea, I.~de~Arriba, I.~Lopez, G.~Hemzal, and A.~Mazzone, ``Remote driving and command of trains: The shift2rail approach.'' \emph{Transportation Research Procedia}, vol.~72, pp. 3723--3729, 2023.

\bibitem{9209333}
M.-P. Pacaux-Lemoine, Q.~Gadmer, and P.~Richard, ``Train remote driving: A human-machine cooperation point of view,'' in \emph{2020 IEEE International Conference on Human-Machine Systems (ICHMS)}, 2020, pp. 1--4.

\bibitem{feng2019toward}
D.~Feng, C.~She, K.~Ying, L.~Lai, Z.~Hou, T.~Q. Quek, Y.~Li, and B.~Vucetic, ``Toward ultrareliable low-latency communications: Typical scenarios, possible solutions, and open issues,'' \emph{IEEE Vehicular Technology Magazine}, vol.~14, no.~2, pp. 94--102, 2019.

\bibitem{chen2018development}
R.~Chen, W.-X. Long, G.~Mao, and C.~Li, ``Development trends of mobile communication systems for railways,'' \emph{IEEE Communications Surveys \& Tutorials}, vol.~20, no.~4, pp. 3131--3141, 2018.

\bibitem{he20225g}
R.~He, B.~Ai, Z.~Zhong, M.~Yang, R.~Chen, J.~Ding, Z.~Ma, G.~Sun, and C.~Liu, ``5g for railways: Next generation railway dedicated communications,'' \emph{IEEE Communications Magazine}, vol.~60, no.~12, pp. 130--136, 2022.

\bibitem{9797698}
Y.~Yu and S.~Lee, ``Remote driving control with real-time video streaming over wireless networks: Design and evaluation,'' \emph{IEEE Access}, vol.~10, pp. 64\,920--64\,932, 2022.

\bibitem{deng2017internet}
L.~Deng, ``Internet engineering task force (ietf) j. xia request for comments: 8286 r. even category: Standards track r. huang,'' 2017.

\bibitem{wiegand2003overview}
T.~Wiegand, G.~J. Sullivan, G.~Bjontegaard, and A.~Luthra, ``Overview of the h. 264/avc video coding standard,'' \emph{IEEE Transactions on circuits and systems for video technology}, vol.~13, no.~7, pp. 560--576, 2003.

\bibitem{wenger2005rfc}
S.~Wenger, M.~Hannuksela, T.~Stockhammer, M.~Westerlund, and D.~Singer, ``Rfc 3984: Rtp payload format for h. 264 video,'' 2005.

\bibitem{Gstreamer}
\BIBentryALTinterwordspacing
Gstreamer: open source multimedia framework. [Online]. Available: \url{https://gstreamer.freedesktop.org/}
\BIBentrySTDinterwordspacing

\end{thebibliography}

\end{document}